# Encryption Algorithm for TCP Session Hijacking

Minghan Chen[1], Fangyan Dai[1], Bingjie Yan[1], Jieren Cheng* and Longjuan Wang

[1]Hainan University, Hainan Province 570228, China

`mh.chen@hainanu.edu.cn,fangyeee@163.com,`
`beiyuouo@foxmail.com, cjr22@163.com*,40552382@qq.com`

**Abstract.** Distributed network of the computer and the design defects of the TCP protocol are given to the network attack to be multiplicative. Based on the simple and open assumptions of the TCP protocol in academic and collaborative communication environments, the protocol lacks secure authentication. In this paper, by adding RSA-based cryptography technology, RSA-based signature technology, DH key exchange algorithm, and HAMC-SHA1 integrity verification technology to the TCP protocol, and propose a security strategy which can effectively defend against TCP session hijacking.

**Keywords:** TCP protocol, session hijacking, security countermeasure, three-time handshake

## 1 Introduction

The transmission of network information is inseparable from the most essential network protocol of the computer. The problem of information transmission affects the size of the cluster and the size of the throughput. The connection-oriented TCP protocol (Transmission Control Protocol) is often used at the transport layer. As the most popular protocol in the world today, the TCP protocol can establish a connection through the three-way handshake method, provide a reliable end-to-end byte transmission stream for network information transmission and prevent packet loss. The purpose of the three-way handshake is to synchronize the serial numbers and acknowledgments of both parties and exchange TCP window size information. However, since protocols such as TCP were formed in the early 1980s, more practical aspects of the protocol were considered during design. Therefore, the TCP protocol data stream is transmitted in plain text, which lacks encryption and authentication of data. There are insufficient considerations in network security, such as: unable to provide reliable identity verification, unable to effectively prevent information leakage, and not providing reliable complete verification of information. And the protocol has no means to occupy and allocate resources, making the TCP protocol fragile and limited. Attacks such as IP spoofing and TCP session hijacking using the TCP protocol are increasing. Therefore, this article studies the problem of TCP plaintext transmission and lack of data encryption. Using the RSA asymmetric encryption algorithm and DH key exchange algorithm algorithm, the existing TCP session process adds identity verification and encryption to ensure a higher algorithm safety. At the same time, this paper improves the single authentication of the original server to the client to achieve double-ended authentication to confirm the authenticity of the data source. In the third handshake process, this article chooses to use HAMC-SHA1 with a hash operation with a key, and performs HMAC-SHA1 integrity verification for each subsequent transmission. Compared with MD5 without a key, In this paper, identity verification is performed while encrypting, which makes the session process more secure and can solve the problems of an attacker disguising as an attacker and communicating with the server.

Compared to other attacks, TCP session hijacking has many benefits, such as sniffing passwords in IP datagrams, especially when using advanced authentication and authentication techniques. However, since all of these advanced authentication techniques occur at the time of the connection, they will not provide protection after this. Therefore, an attacker can enter the system simply by hijacking a legitimate connection. This allows an attacker to appear as a legitimate user in the operating system security mechanism [1].

From 90 mid-year introduction of cookies since the use of cookies loopholes session hijacking attacks has been a security problem. In the article, Vissiago [2] describes the use of Web application design flaws ED for session management attacks and discusses the "current prevalence" of these defects. This survey emphasizes that cookies are vulnerable to session hijacking and are not a source of ability to maintain web authentication. Because of this, many researchers began to study network security protocols. Wannes Meert and some other people in 2001 publish a paper "SessionShield:Lightweight Protection Against the Session Hijacking" [3]which presented SessionShield - a lightweight client protection mechanisms. Prevent session hijacking by using session identifier values not used by legitimate client scripts.Dacosta and some other researcher [4] proposed a more robust session authentication scheme for one-time cookies (OTC) as a solution for session hijacking. Each user request is signed by using a session secret that is securely stored in the browser. Each token is only allowed to be used once. One cookie can significantly improve the security of the web application with minimal impact on performance.

However, cookies have a serious design flaw that limits their security. In particular, cookies cannot provide session integrity for an attacker capable of hosting content on the relevant domain. Therefore, A.Bortz [5] proposed a new concept of related domains that are vulnerable to session hijacking attacks. They proposed a solution to suppress attacks and lightweight expansion of cookies which are secure against related domains and network attackers. By this way, they result in session integrity.

The simplest and most effective defense against IP spoofing, TCP spoofing and TCP session hijacking are those that provide Inteet access. If all of these organizations have sufficient responsibility to prevent IP datagrams coming from outside the network from reaching the network, the above attacks cannot be performed. Unfortunately, many networks offer unregulated Internet access. Therefore, other means of preventing fraud and hijacking attacks must be used. The easiest and most effective way for an organization is to block all IP datagrams from the network, and a properly configured firewall can be used to enforce such policies. Therefore，Elie [6] exploited the features of a web browser to develop a secure login system-Session Juggler, from an untrusted terminal and provided a secure logout mechanism. In 2012, Asif and some researchers [7] found that the efficient identity management system OpenID is easily hijacked by the session. In order to solve the theft of user identity information caused by session hijacking, ELie and other researchers proposed a two-factor authentication method. It authenticates the user by checking the credentials and PIN code stored on the ID server. Even if the session between the user and the identity provider is hijacked by an intruder, the intruder may not be able to access the PIN due to the two-factor authentication system. Burgers and other researchers [8] made a new way to prevent theft of the session, via the secure communication channel negotiation bound to the application user authentication, a server is introduced independently of the client and server software running a reverse proxy, the mechanism Established on the communication channel of security negotiation. This is achieved by establishing a server-side reverse proxy. It runs independently of the client and server software.

It is also important to assess threats. For example, an attacker is less likely to encounter hijacking an anonymous FTP session. In this regard, Stango and other researchers [9] used threat analysis to show how potential opponents can use system weaknesses to achieve their goals. It identifies threats and defines risk mitigation strategies for specific architectures,features and configurations. Desmet [10] analyzed the threats that might be associated with using web services in web applications. In session hijacking, the intruder mimics the identity of the victim and uses the same resource access rights as the victim. The consequences can be catastrophic because it can lead to the loss of critical information. Therefore, session hijacking has always been the focus of researchers, and they have proposed strategies to prevent and mitigate session hijacking.

## 2      Background and relevant work

### 2.1      Attack on TCP

**SYN attack.** A SYN attack is a denial of service attack against the three-way handshake of the TCP protocol. In the server receives from the client SYN after a request packet will be responded by ACK, and reserve resources for the client until the client from listening to the ACK packet start communicating parties. If the client sends a large number of SYN packets in a short interval, the limited resources of the server will be exhausted, the server will not continue to work, and the network hacker successfully implements the denial of service attack. (complementary map)
**Session Hijacking.** The TCP protocol stipulates that authentication between two hosts takes place during the connection establishment phase (Handshake stage), after which no authentication is required. The lack of security for TCP session authentication gives network hackers a chance. After the client establishes a connection with the server, the attacker can send a fake RST message to the client, causing the client to terminate the session connection with the server. At this point, the network hacker forges the data packet to communicate with the server based on the obtained packet identifier (the client's IP address, the client's port, the server's IP address, the server's port, the serial number, and the acknowledgment number).

## 3 Defense mechanism for TCP session hijacking

### 3.1 Improved TCP handshake scheme

**Problem solving process.**
*The first handshake.* Client by client and server issuing common certification authority to a third party server obtaining the digital certificate server public key S+; client to select a large prime number P and P is a primitive root of A, and the selected private XC(XC<P); client according DH secret key exchange algorithm and P, a computing client public key YC; client with RSA encryption algorithm to S+ is a secret key of large prime numbers P||A||YC|| bits Description the combined data is encrypted ciphertext C. client will SYN set a randomly selected sequence number seq is J, the SYN packet header splicing C sent to the Server.

*Second handshake.* After server receiving the data packet, the data portion client and server mutual recognition of third party certification authority's private key S- to C for RSA decryption algorithm, to obtain a large prime number P, the original root A and YC ; server selected private key XS(XS<P); the DH exchange algorithm and key P, a calculation server public key YS; the DH key exchange algorithm and YC, P calculate the session key K; server to client, and server third-party certification authority's private key common authentication S- as a key in accordance with RSA algorithm, YS signing obtain S; server will SYN set to 1, the ACK is set 1, randomly selected sequence number seq is I, the acknowledgment number ack For J, the ACK message header is spliced as data S to send the data packet to the client.

*The third handshake.* After the client receives the data packet, it first checks whether the ACK is 1, and confirms whether the ack is J. If not, it discards it. If it is, it uses S+ as the key and S is the message input. According to the RSA algorithm, it is obtained. The server 's public key YS; client uses the YS and DH key exchange algorithm P, to calculate the session key K; the client uses the session key K as the key, according to the HAMC-SHA1 algorithm to generate the message hash value HAMC; client will ACK Set to 1, ack is I, with the HAMC as the data, send the packet to the server.

After receiving the data packet, the server first checks whether the ACK is 1, and whether ack is I. If it is not discarded, if yes, the session key K is used as the key, and the HAMC' is calculated according to the HAMC-SHA1 algorithm. If HAMC' =HAMC, Then the connection is successfully established. 4. In the data transmission phase, every 100 data packet senders use the session key K as a key, and according to the HAMC-SHA1 algorithm, calculate the HAMC; the sender splicing data packets are sent together with the HAMC;

After receiving the data packet, the receiver uses the session key K as the key and calculates the HAMC' according to the HAMC-SHA1 algorithm. Only when HAMC=HAMC', the transmission can continue.

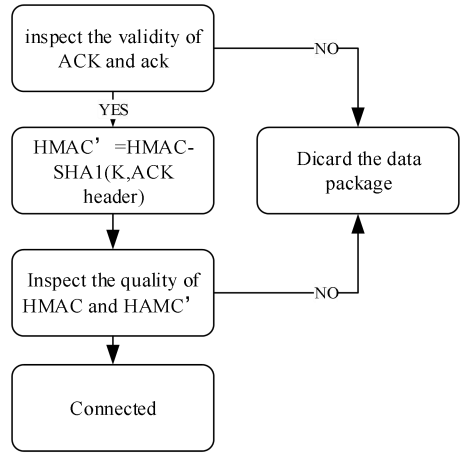

**Fig. 1.** Third shake of server

## 4  Experiment analysis

### 4.1  TCP session hijacking experiment

**TCP Session hijacking experiment process.** The server creates a socket object, binds the IP address and port number (192.168.0.105, 49999), and then listen on the 49999 port number. The client creates socket object, binds IP address and port number (192.168.0.104, 59999), then send a connection request to the server (192.168.0.105 49999). The server receives the client's connection request and creates a thread to handle the connection verification of the client's TCP protocol. The server and client perform a TCP protocol three-way handshake to verify identity. The attacker use Ettercap to implement ARP attack to make the client mistakenly think it is a gateway and send packets to the attacker. Then the attacker capture the packets that are transferred between the server and the client, as shown in the picture below. The attacker uses the information in the TCP packet. The required information is as follows: using TCP protocol; Time To Live: 128; Source IP address: 192.168.0.104; Destination IP address: 192.168.0.105; Source Port: 59999; Destination Port: 49999; Next Sequence number: 2775375568; Acknowledgment number: 2356415172.

The attacker uses the Spoof Ip4Tcp packet option in the Netwag tool and enter the command "sudo passwd root" that attacker wants to send to the server to generate the Netwox command.

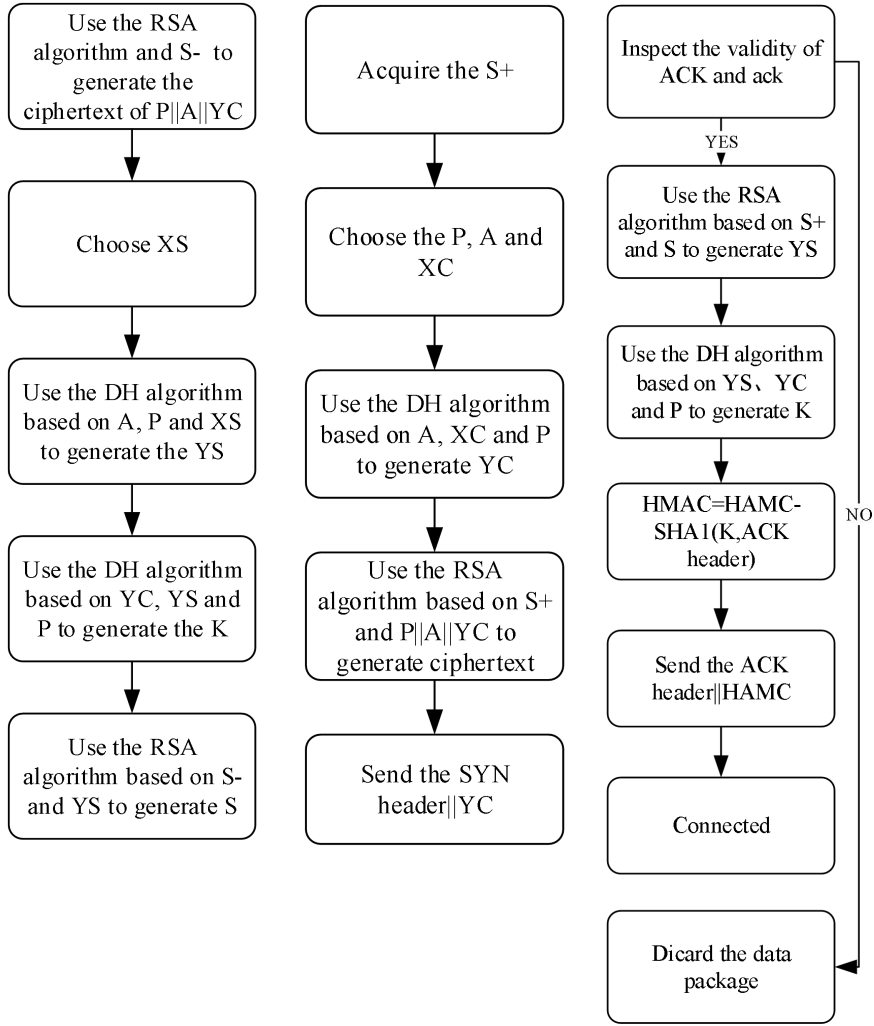

**Fig. 2.** First shake of client     **Fig. 3.** Second shake of server     **Fig. 4.** Third shake of client

```
40 --ip4-ttl 128 --ip4-protocol 6 --ip4-src 192.168.0.104 --ip4-dst 192.168.0.105 --
tcp-src 59999 --tcp-dst 49999 --tcp-seqnum 2775375568 --tcp -acknum
2356415172 --tcp-ack --tcp-psh --tcp-window 128 --tcp-data "'sudo passwd root'"
```

The attacker uses the Netwox tool to generate and send fake TCP packets to the server.

The server receives the data packet sent by the attacker. After verification, it is mistaken that it is the data packet sent by the client, correctly receives the data packet. and execute the commands. The output on the server side is as follows.

> Listening…
> Accept!
> Connection addr: (192.168.0.104, 59999)
> Recv: sudo passwd root

You can see that the server received the packet forged by the attacker and thought that the packet came from the client (192.168.0.104: 59999). Then the attacker forges the server to send the packet containing RST command to the client to let the client go offline. After that, the attacker gets all the permissions of the client on the server.

**Encryption experiment process.** Note: For the convenience of experiment, the number of prime p used in the DH key exchange protocol used in the experiment is small, but it can be recommended according to RPC 3626 in the actual environment should use a larger prime number, so that the attacker can not hack the server's or client private's key in the DH key exchange protocol in a limited time.

**Specific experimental process.** In the first handshake session, the client obtains the server's public key S+ from the digital certificate issued by the third-party authority authenticated by the client and the server to the server:

> S+:(728947908706520468702710835614097609778934317790345395720490862
> 5740114208903632340542215628999136015415388380418420294863244335815
> 571945495531508812116349,65537)

and selects the prime number p=97, the original root A=5, the private key XC=36 in the DH key exchange protocol, calculate the public key $YC=A^{XA}(\mod p)=50$ in the DH key exchange protocol. The information is stitched in the order of the order negotiated with the server 'p$A$XA$', and the data is obtained by encrypting with the RSA encryption algorithm:

> b'v\xbf\x8f\x85+\x9f\xb4\xcd|\x17\x81\xe9y\xcd\x96\xa6\xd8\xbe\x80\x89\x04l\
> x0e\xb6\\\xc2y\x93iO\xb0sBL\x84D\x13\xf33E]\xf8\xe4\x97\t\xb1\xb0\xb4\xb5\
> x0e\xfd9\xa0]\xf6\x19\xb7\xedh\xf6\x84G\x8e'

At the same time, the client sets SYN to 1, randomly selects the serial number seq as J, and packs it with other information into a TCP packet and sends it to the server.

The server receives the TCP packet sent by the client, and uses the RSA private key corresponding to the public key when the digital certificate issued by the client and the server co-certified by the server and the server authenticates the public key:

> S-:(728947908706520468702710835614097609778934317790345395720490862
> 5740114208903632340542215628999136015415388380418420294863244335815
> 5719454955315088121163 49,65537)

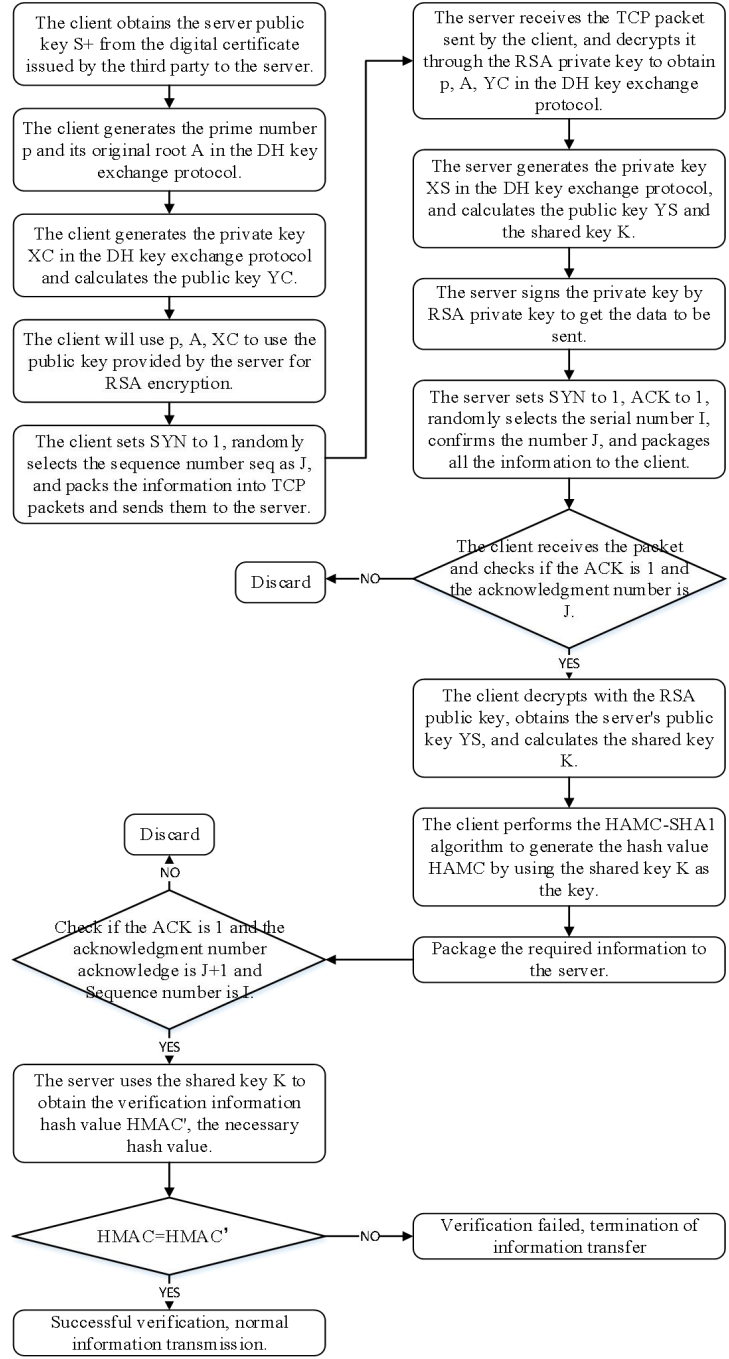

**Fig. 5.** improved three handshakes of TCP

Then the server decrypts the information sent by the client, and the information can only be decrypted by the server. After that the server can obtain the prime number p, the original root A and the client public key YC in the DH key exchange protocol. The server selects the private key XS=58 in the DH key exchange protocol, calculates the public key $YS=A^{XB} \pmod{p}=44$ in the DH key exchange protocol, and then calculates the shared key $K=YC^{XS} \bmod p = 50^{58} \bmod 97 = 75$ in the DH key exchange protocol.

The server encrypts the public key YS in its own DH key exchange protocol by its own RSA private key to obtain the data:

```
b'e\xe3J\xbdF\x93^\x86\xa9\xba\xe3d\xc6\x94R\xd3_\xaf\xb1\xefL_%\x8a\x0e\
xdb\x19\xc7\x8a>\x8c\xea\x8d\xf1)P\xe0q\xa5\t\xa5\xa9+{\x8fd0<\x8b\x19b\xbe\
x86\xb0\x86yf\xcd\xa0@\xf7\xe8C3'
```

The server sets SYN to 1, ACK to 1, randomly selects the sequence number seq as I, confirms the acknowledgment number as J, and packs it and other required information into a TCP packet and sends them to the client.

After the client receives the TCP packet sent to him by the server, it first checks that the ACK is 1, the confirmation acknowledgment number is J. Then, the data packet is decrypted by using the RSA public key S+ on the server digital certificate as the key, and the data the public key YS in the server DH key exchange protocol contained in the data packet is obtained.

The client then calculates the shared key K using the DH key exchange protocol YS and according to $K=XC^{YS}=50^{58} \bmod 97=75$. The client generates a mutual authentication message='message' hash value :

$$HAMC=ada668f4688e906e157d8613dc4408ce00de1cf0$$

According to the HAMC-SHA1 algorithm using the shared key K in the DH key exchange protocol as the key.

The client sets ACK to 1 and acknowledgment number to I, which packs the HAMC and other TCP packet headers and sends them to the server.

After receiving the data packet, the server first checks that the ACK is 1, and the acknowledgment is I. Then, using the shared key K=75 as the secret key.

$$HAMC'=ada668f4688e906e157d8613dc4408ce00de1cf0$$

It is calculated according to the HAMC-SHA1 algorithm to verify whether

$$HAMC'=HAMC.$$

After the two parties successfully establish a connection, it is required to perform verification once every 100 data packets. The K=75 in the DH key exchange protocol is used as the key. According to the HAMC-SHA1 algorithm, the HAMC is calculated, and the data to be sent is sent together with the HAMC data, and the party that accepts the verification uses the DH key exchange protocol in its own. K=75, calculate HAMC' for verification. Message transmission can continue if and only if HAMC = HAMC'.

**Experimental results.** After experiment, this paper used the modified encrypted TCP protocol verification, because it is impossible to brute force all possible private keys in the theoretical time complexity, and both parties will perform message verification every time period, which can prevent the attacker from pairing the client. The session hijacking of the server cannot obtain the operation permission of the client to the server, and the injection operation of the attacker to the server is extremely effectively avoided.

## 5 Discuss

### 5.1 Limitations

The defense mechanism based on TCP session hijacking proposed in this paper can effectively solve the session hijacking that occurs during the TCP three-way handshake phase. However, in the data transmission process, the HAMC-SHA1 signature and integrity verification for each data packet will spend a lot of system resources.

### 5.2 Solution

In order to ensure integrity verification and reduce overhead, we currently propose a window value of 100. When the number of packets sent by the client reaches 100, the 100th data packet is signed and the counter is cleared to 0. After receiving the data packet with the HAMC signature, the server performs integrity verification according to the negotiated HAMC-SHA1 algorithm and the session key K. Therefore, it is necessary to obtain a reasonable window value through a large number of experimental and scientific mathematical proofs, and achieve the best balance between resource consumption and security.

### 5.3 Advantages

**HAMC-SHA1 Algorithm.** Currently used integrity algorithms are MD5, MD4, SHA1 and others. SHA1 is one of the most widely used algorithms. HMAC is a key-related hash operation message authentication code. The HMAC algorithm using hashing algorithm and using a message and a key as input, generates a message digest as output. We use the HAMC-SHA1 algorithm to use the session key as the key to perform identity verification and complete the authentication.
**Window value based integrity verification.** To completely prevent the session be hijacked, it is the most reliable method for verifying the integrity of each data packet based on the HAMC-SHA1 and session secret key. Therefore we have proposed a more scientific window value. When the number of transmitted packets reaches the window value, both parties perform integrity verification and the counter is reset to 0. It can find a relative balance between security and resource consumption.

# 6 Conclusion

The design flaw of the TCP protocol is an objective cause of hacker attacks, and there are many technologies that can protect the network from attacks at present. This paper proposes a security mechanism based on TCP three-way handshake to defend against a well-known attack method - TCP session hijacking. First we show how to implement session hijacking, and then we add cryptography and authentication techniques to the TCP protocol for session hijacking. Through the three-way handshake phase of TCP protocol combined with cryptography to enhance identity verification, we propose a security countermeasure for session hijacking based on TCP protocol. This security countermeasure not only can effectively identify malicious packets constructed by network attackers, but also protect the protocol security. At the same time, it can also sign the authentication label of the data packet (client IP address, client port, server IP address, server port, serial number and confirmation number). Once these identifiers are changed, they are considered to be fake packets constructed by the attacker. If the packets can not pass the integrity verification on the server side, the server will discard the packets. The server only processes the data in the packet after the integrity verification is passed, and the integrity verification is based on the HAMC-SHA1 algorithm, where the secret key is the session key of the first handshake and the second handshake. In this way, in the case of relatively small resource consumption, integrity verification and authentication are implemented, and TCP session hijacking attacks are effectively blocked.

Future work includes a large number of cyberattack experiments and mathematical proofs, taking reasonable window values and find a balance between security and resource consumption. When the window value is smaller, the success rate of successfully intercepting TCP session hijacking is higher, but the resource consumption of the server and client will become very large; when the window value is larger, it is possible for a network attacker to take control of the communication details of the packet before successful hijacking until the next integrity check, and the security is reduced. Therefore, finding a scientific and reasonable window value through mathematical proof and a large number of experimental tests is the key to the practical application of this strategy.